\documentclass[aps,pra,twocolumn,superscriptaddress,a4paper,noeprint]{revtex4-2}
\usepackage{graphicx}

\usepackage{amsmath}
\usepackage{float}
\usepackage{color}
\usepackage{textcomp}
\usepackage{upgreek}
\usepackage{amsfonts}

\pdfoutput=1 

\usepackage{fancyhdr}
\usepackage[colorlinks=true,citecolor=blue,linkcolor=magenta]{hyperref}
\hypersetup{
pdfauthor = {},
colorlinks = true, linkcolor = blue, urlcolor=blue, bookmarksnumbered =  true}

\begin{document}
\title{High quality-factor diamond-confined open microcavity}
\author{Sigurd Fl\aa gan}
\email{sigurd.flagan@unibas.ch}
\affiliation{Department of Physics, University of Basel, Klingelbergstrasse 82, Basel CH-4056, Switzerland}
\author{Daniel Riedel}
\altaffiliation{Current address: E. L. Ginzton Laboratory, Stanford University, Stanford, CA 94305, USA}
\affiliation{Department of Physics, University of Basel, Klingelbergstrasse 82, Basel CH-4056, Switzerland}
\author{Alisa Javadi}
\affiliation{Department of Physics, University of Basel, Klingelbergstrasse 82, Basel CH-4056, Switzerland}
\author{Tomasz Jakubczyk }
\altaffiliation{Current address: Faculty of Physics, University of Warsaw, 02-093 Warsaw, Poland}
\affiliation{Department of Physics, University of Basel, Klingelbergstrasse 82, Basel CH-4056, Switzerland}
\author{Patrick Maletinsky}
\affiliation{Department of Physics, University of Basel, Klingelbergstrasse 82, Basel CH-4056, Switzerland}
\author{Richard J. Warburton}
\affiliation{Department of Physics, University of Basel, Klingelbergstrasse 82, Basel CH-4056, Switzerland}
\date{\today}

\begin{abstract}
With a highly coherent, optically addressable electron spin, the nitrogen vacancy (NV) centre in diamond is a promising candidate for a node in a quantum network. However, the NV centre is a poor source of coherent single photons owing to a long radiative lifetime, a small branching ratio into the zero-phonon line (ZPL) and a poor extraction efficiency out of the high-index host material. In principle, these three shortcomings can be addressed by resonant coupling to a single mode of an optical cavity. Utilising the weak-coupling regime of cavity electrodynamics, resonant coupling between the ZPL and a single cavity-mode enhances the transition rate and branching ratio into the ZPL. Furthermore, the cavity channels the light into a well-defined mode thereby facilitating detection with external optics. Here, we present an open Fabry-Perot microcavity geometry containing a single-crystal diamond membrane, which operates in a regime where the vacuum electric field is strongly confined to the diamond membrane. There is a field anti-node at the diamond-air interface. Despite the presence of surface losses, quality factors exceeding $120\,000$ and a finesse  $\mathcal{F}=11\,500$ were observed. We investigate the interplay between different loss mechanisms, and the impact these loss channels have on the performance of the cavity. This analysis suggests that the ``waviness" (roughness with a spatial frequency comparable to that of the microcavity mode) is the mechanism preventing the quality factors from reaching even higher values. Finally, we apply the extracted cavity parameters to the NV centre and calculate a predicted Purcell factor exceeding 150.
\end{abstract}

\maketitle
\section{Introduction}
The development of an efficient interface between stationary and flying qubits\,\cite{Gao2015,Atature2018} is an essential step towards the realisation of large-scale distributed quantum networks\,\cite{Duan2010,Simon2017} and the quantum internet\,\cite{Kimble2008,Wehner2018}. In such a network, quantum nodes with the ability to store and process quantum information are interconnected via quantum channels in order to distribute quantum information and entanglement across the network\,\cite{Ritter2012,Reiserer2015}. To communicate between remote network nodes, optical photons are a convenient choice owing to low absorption and decoherence\,\cite{Simon2017,Northup2014}, alongside compatibility with pre-existing classical fibre-networks\,\cite{Reiserer2015,Valivarthi2016,Yu2020}. However, transmission of quantum information over long distances remains a challenge owing to photon propagation-loss in the network links\,\cite{Sangouard2011,Yin2017}.

Quantum-repeater protocols represent a means to compensate for photon-loss\,\cite{Briegel1998}. In principle, entanglement can be distributed over long distances by pair-wise entanglement swapping of adjacent nodes, where each network link covers a sub-section of the total distance\,\cite{Duan2001,Sangouard2011}. These network nodes require high-fidelity processing of quantum information combined with a robust, long-lived quantum memory\,\cite{Munro2015,Reiserer2015,Bhaskar2020}. Long-lived, optically addressable spins in the solid-state have emerged as a promising candidate\,\cite{Gao2015,Awschalom2018,Atature2018}. The development of an efficient spin-photon interface\,\cite{Borregaard2019} is limited by the weak cross-section between single spins and photons\,\cite{Reiserer2015}. Crucially, this interaction can be enhanced by embedding the solid-state spins inside optical resonators\,\cite{Vahala2003,Reiserer2015,Janitz2020}.

Owing to its highly coherent\,\cite{Balasubramanian2009,Bar-Gill2013}, optically addressable electron-spin\,\cite{Robledo2011,Chu2015,Irber2021} and the possibility of coherent couplings to nearby nuclear spins\,\cite{Maurer2012,Taminiau2014,Yang2016,Abobeih2018,Bradley2019}, the negatively charged nitrogen-vacancy (NV) centre in diamond is a promising candidate as a stationary qubit in a quantum network\,\cite{Childress2005,Childress2006}. Advances in spin-photon\,\cite{Togan2010} and spin-spin entanglement\,\cite{Bernien2013} have paved the way for long-distance entanglement\,\cite{Hensen2015}, quantum teleportation\,\cite{Pfaff2014}, entanglement distillation\,\cite{Kalb2017} and on-demand entanglement delivery\,\cite{Humphreys2018}, all key steps towards the realisation of a quantum network\,\cite{Pompili2021}. However, the scalability of these experiments is limited by the modest entanglement rates, in turn limited by the small flux of coherent photons\,\cite{Riedel2017}.

For NV centres in diamond, the generation rate of coherent photons is limited by the long radiative lifetime ($\tau_{\textrm{0}}\simeq 12\,\textrm{ns}$) and the small branching-ratio ($\sim 3\%$) of photons into the zero-phonon line (ZPL)\,\cite{Barclay2011}. Furthermore, the photon extraction efficiency out of the diamond is poor owing to total internal reflection at the diamond-air interface ($n_{\textrm{d}}=2.41$). In principle, these problems can be addressed by resonant coupling of the ZPL emission to photonic resonators with a high ratio of quality factor ($\mathcal{Q}$) to mode volume $V$\,\cite{Su2008,Riedel2017,Ruf2021}. The cavity enhances the ZPL emission on two grounds. First, the cavity provides a well-defined output mode, ideally a Gaussian, leading to improved detection efficiency\,\cite{Gao2015,Riedel2020}. Secondly, utilising the Purcell effect\,\cite{Purcell1946}, a cavity resonant with the ZPL enhances the total transition rate and likewise the proportion of the photons emitted into the ZPL\,\cite{Riedel2017}.

Resonant enhancement of the ZPL has been demonstrated in photonic crystal cavities\,\cite{Faraon2012,Riedrich-Moller2015,Jung2019}, hybrid-\,\cite{Barclay2011,Gould2016,Schmidgall2018} and microring resonators\,\cite{Faraon2011}. While these resonators offer a large Purcell factor, the NV centres suffer from poor optical coherence, compromising the photon indistinguishability. This inhomogeneous broadening of the ZPL is a consequence of a fluctuating charge environment presumably caused by fabrication-induced surface damage\,\cite{Riedel2017,Ruf2021}. Increasing the quality of the crystalline environment has proven to be beneficial\,\cite{Ruf2019,Kasperczyk2020,Yurgens2021}.

Open Fabry-Perot microcavities offer an alternative to photonic crystal cavities. The required fabrication is relatively modest: only micron-sized single-crystalline membranes of the host material are required. A reasonably small mode-volume and a high $\mathcal{Q}$-factor can be achieved. Furthermore, the Fabry-Perot cavity offers full \textit{in situ} spatial and spectral tunability along with a Gaussian output mode\,\cite{Barbour2011,Riedel2017,Riedel2020}. As a consequence, open Fabry-Perot cavities offer an attractive platform to enhance the emission from various single-photon emitters embedded in solid-state hosts\,\cite{Greuter2015,Riedel2017,Wang2017a,Wang2019,Casabone2020,Jensen2020,Tomm2021,Merkel2020,Haussler2020,Ruf2021}. 

\begin{figure}
\includegraphics[width=0.48\textwidth]{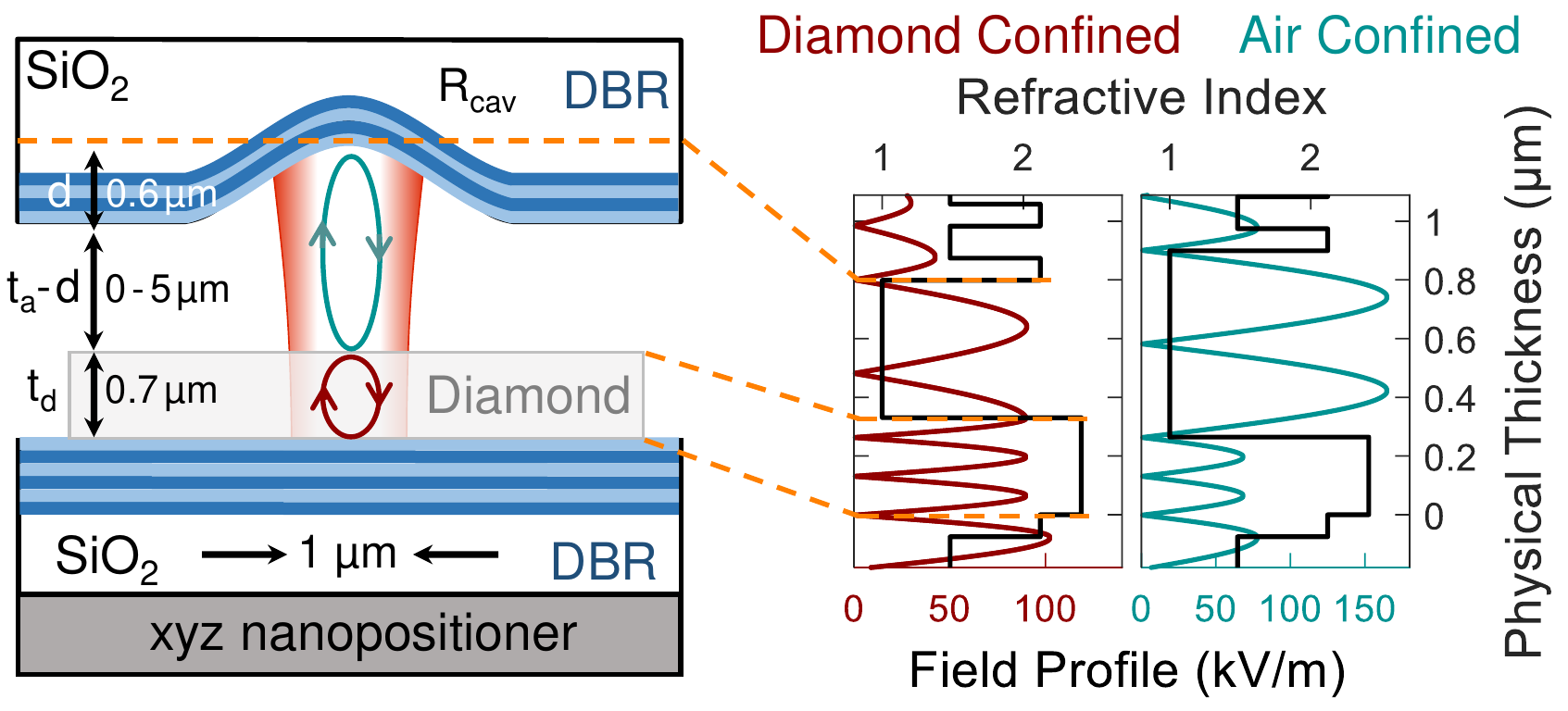}
\caption{Schematic of the diamond membrane embedded into an open Fabry-Perot cavity. In the diamond-confined regime, the vacuum electric field is strongly confined to the diamond.  Furthermore, the field profile possesses an anti-node across the diamond-air interface. In the air-confined regime, there is a field node across the diamond-air interface, and the vacuum electric field is strongly confined to the air-gap.}
\label{fig:Cav}
\end{figure}

In this work, we present a diamond membrane embedded in a Fabry-Perot microcavity operating in the so-called ``diamond-confined" regime\,\cite{Janitz2015,Jensen2020}. In this regime, there is a vacuum-field anti-node at the diamond-air interface -- the design is prone to scattering losses at this interface (Fig.\,\ref{fig:Cav}). Despite this loss channel, $\mathcal{Q}$-factors of more than $10^5$ were observed for short (few $\lambda$) cavity lengths. The measured $\mathcal{Q}$-factor along with the low scattering-cross-section at the diamond surface lead us to predict a Purcell factor greater than 150 for the ZPL. Although the motivation behind this work is to enhance the flux of coherent photons from NV centres in diamond, the theoretical Purcell-factor depends solely on the cavity parameters. Therefore, similar results would be expected for other defect centres in crystalline hosts provided the surface losses are reduced sufficiently. 

\section{Methods}
\begin{figure}
\includegraphics[width=0.48\textwidth]{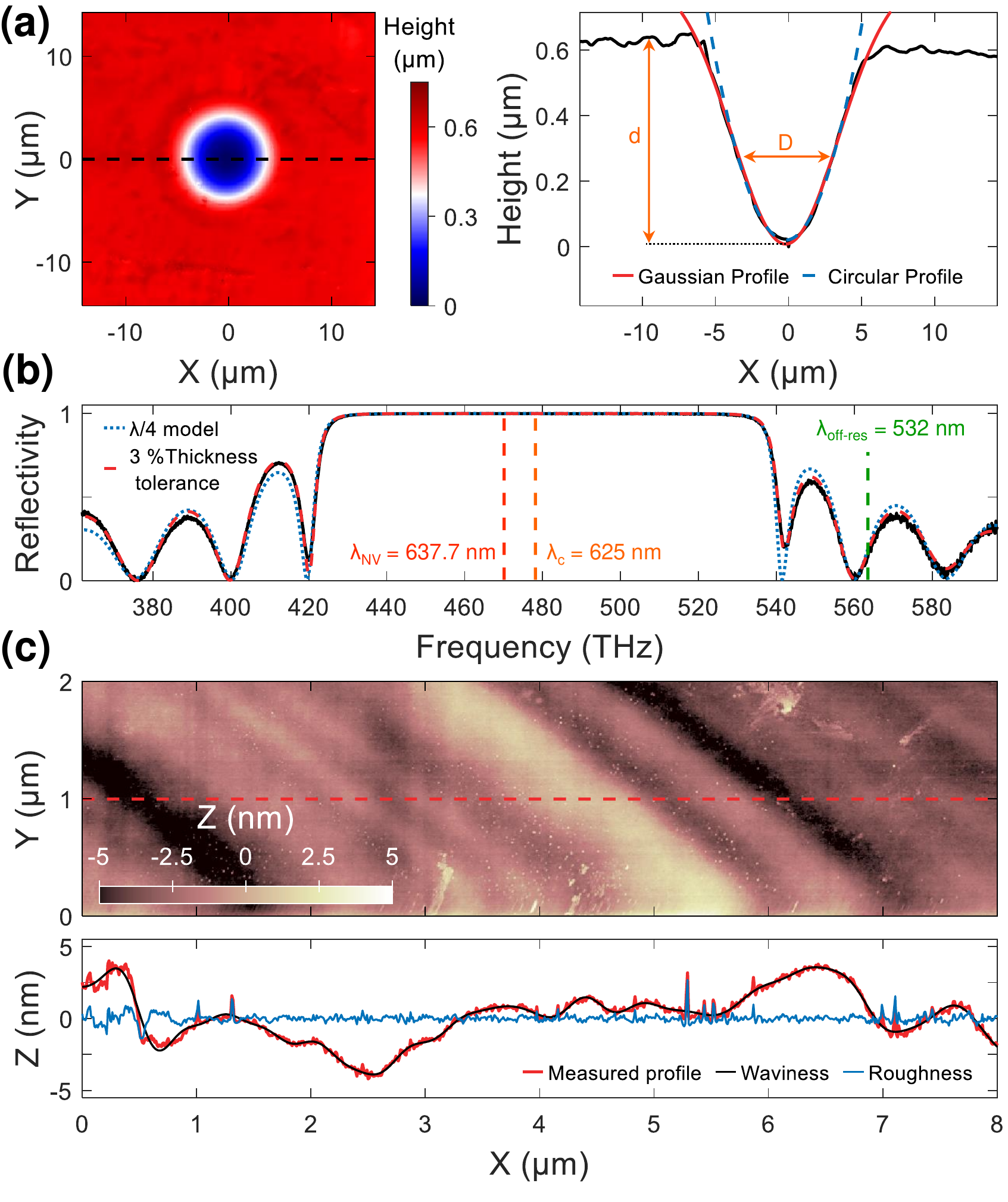}
\caption{\textbf{(a)} The left panel shows a laser-scanning confocal-microscope image of the crater used in this experiment. The geometrical parameters of the cavity were extracted by analysing the surface profile along the axis of the crater (right panel). The radius $R_{\textrm{cav}}= (19.7\pm 2.5)\,\upmu\textrm{m}$ and crater depth  $d = 0.64\,\upmu\textrm{m}$ were extracted from a Gaussian fit (Eq.\,\ref{eq:MirrorProfile}). \textbf{(b)} Transmission measurement of the DBR mirror using a white-light source normalised to the transmission through an uncoated SiO$_{\textrm{2}}$ chip. By fitting the reflectivity spectrum using a transfer-matrix based refinement algorithm, the stopband centre was determined to lie at $\lambda_{\textrm{c}}=625\,\textrm{nm}$. \textbf{(c)} The top panel shows an AFM measurement of the diamond membrane bonded to the DBR. Large-range structures attributed to polishing marks are visible. The bottom panel shows the surface profile along the line-cut marked in red. From the black line, a surface waviness $W_{\textrm{q}}=1.6\,\textrm{nm}$ was extracted. The blue line shows the residual short-range surface-roughness with RMS roughness of $\sigma_{\textrm{q}} = 0.3\,\textrm{nm}$.  }
\label{fig:intro}
\end{figure}

The device used in this work is a highly miniaturised planar-concave Fabry-Perot cavity, depicted schematically in Fig.\,\ref{fig:Cav}. The cavity mirrors are created from a SiO$_2$ substrate, where, for the top mirror, a CO$_2$-laser ablation-technique was used to create atomically smooth craters with a radius-of-curvature $R_{\textrm{cav}}\sim 10\dots30\,\upmu\textrm{m}$\,\cite{Hunger2012,Najer2017}. The profile of the crater was determined using a laser-scanning confocal-microscope image (Keyence Corporation, resolution  $\sim 200\,\textrm{nm}$), as displayed in Fig.\,\ref{fig:intro}\,(a). The surface profile of the radial cross section of the curved mirror can be described by 
\begin{equation}
z(r)=-d\exp\left(-\frac{r^2}{2R_{\textrm{cav}}d}\right)\,. \label{eq:MirrorProfile}
\end{equation}
Fitting a truncated Gaussian (Eq.\,\ref{eq:MirrorProfile}) to the surface profile yields ${R_{\textrm{cav}}=(19.7\pm 2.5)\,\upmu\textrm{m}}$ and a depth $d=0.64\,\upmu\textrm{m}$. By comparison, a circular fit to the crater yields ${R_{\textrm{cav}}=21\,\upmu\textrm{m}}$. 

After fabrication, the mirror substrates were coated with a high-reflectivity distributed-Bragg-reflector (DBR) coating (ECI evapcoat), consisting of 14 (15) $\lambda/4$ layers of SiO$_2$ ($n=1.46$) and Ta$_2$O$_5$ ($n=2.11$) for the top (bottom) mirror, respectively. The reflective coatings were characterised using the transmission from a white-light source, normalised to the transmission of an uncoated SiO$_2$ substrate (Fig.\,\ref{fig:intro}\,(b))\,\cite{Riedel2020}. Using a transfer-matrix-based calculation (Essential Macleod) we were able to reconstruct the reflective stopband based on a $\frac{\lambda}{4}$ model (blue line Fig.\,\ref{fig:intro}\,(b)). By further allowing for a 3\,\% tolerance on each individual layer thickness, the exact mirror structure could be reconstructed (red line Fig.\,\ref{fig:intro}\,(b)). From this calculation, we deduce a stopband centre of $\lambda_{\textrm{c}} = 625\,\textrm{nm}$.

Following previously reported fabrication procedures, a diamond micro-membrane of typical dimensions $\sim35\times35\times0.7\,\upmu\textrm{m}^3$ was fabricated from commercially available single-crystalline diamond\,\cite{Maletinsky2012,Appel2016,Riedel2014}. Post fabrication, the diamond membrane was transferred to the bottom DBR using a micro-manipulator. The small contact area, combined with a low surface roughness, facilitates bonding of the diamond membrane to the bottom mirror via van der Waals interactions\,\cite{Riedel2017,Riedel2020}. After transfer, the surface quality of the top-surface of the diamond membrane was investigated by atomic-force microscopy (AFM) (Fig.\,\ref{fig:intro}\,(c)). Analysing the AFM image revealed a large-scale surface texture  (period $\sim \upmu\textrm{m}$) with root-mean-square (RMS) waviness of vertical height $W_{\textrm{q}}=1.6\,\textrm{nm}$, which we attribute to polishing marks, and a small-scale RMS surface-roughness of $\sigma_{\textrm{q}}=0.3\,\textrm{nm}$ (period sub-nm).

After characterisation of the DBRs and the diamond membrane, the bottom mirror was attached to the top-surface of a three-axis piezo-electric nano-positioner (attocube), and the entire piece then mounted inside a homebuilt titanium cage. The top mirror was glued onto a titanium holder; the holder was attached to the top of the cage with a thin layer of indium between holder and cage. The soft indium acts as an adjustable spacer allowing the relative tilt between the two mirrors to be minimised. The piezo-electric positioners allow the microcavity length and lateral position of the microcavity mode to be adjusted {\em in situ} \,\cite{Greuter2014,Riedel2020}. Although all measurements in this work were carried out at room temperature, the compact design facilitates experiments in a 4\,K liquid-helium bath-cryostat\,\cite{Barbour2011,Greuter2015,Riedel2017,Najer2019,Tomm2021}.\\

As a first characterisation, we aim to extract the geometrical parameters of the cavity by using a simple model based on Gaussian optics\,\cite{Greuter2014}. The radius-of-curvature, $R_{\textrm{cav}}$, of the curved mirror can be extracted by analysing the spacing between the fundamental $(q,0,0)$ and higher-order $(q,n,m)$ modes. The cavity length, $L_{\textrm{eff}}$, the mode number $(q,n,m)$ and $R_{\textrm{cav}}$ are linked via\,\cite{Greuter2014,Riedel2020}
\begin{equation}
L_{\textrm{eff}}(q,n,m)=\left[q+\frac{n+m+1}{\pi}\cos^{-1}\left(\sqrt{g}\right)\right]\frac{\lambda_0}{2}\,, \label{Effective_Cavity_Length}
\end{equation}
where $g=1-\frac{L_{\textrm{eff}}(q,n,m)}{R_{\textrm{cav}}}$. Here, the effective cavity length $L_{\textrm{eff}}$ is defined as the physical separation between the two mirrors , the air-gap, plus the field penetration depth into each mirror upon reflection\,\cite{Riedel2020,Koks2021}.

To put photons into the cavity mode, we rely on the diamond as an internal light source\,\cite{Riedel2020}. We pump the diamond with a green continuous-wave laser (Laser Quantum Ventus532, $\lambda = 532\,\text{nm}$, $P = 30\,\text{mW}$) whose wavelength lies on the blue-side of the stopband of the DBRs (Fig.\,\ref{fig:intro}\,(b)). We collect the resulting photoluminescence (PL), here background PL from the diamond, while stepwise reducing the width of the air-gap $t_{\textrm{a}}$ by applying a positive voltage to the z-piezo. Working in a backscattering geometry, the PL signal is coupled into a single-mode fibre (Thorlabs 630HP) and then sent to a spectrometer (Princeton Instruments). A long-pass filter (Semrock LP03-532RS-25) and a dichroic mirror (Semrock, FF560-FDi01) are used to filter out the excitation laser from the PL signal\,\cite{Riedel2020}. 

\begin{figure}[t]
\includegraphics[width=0.48\textwidth]{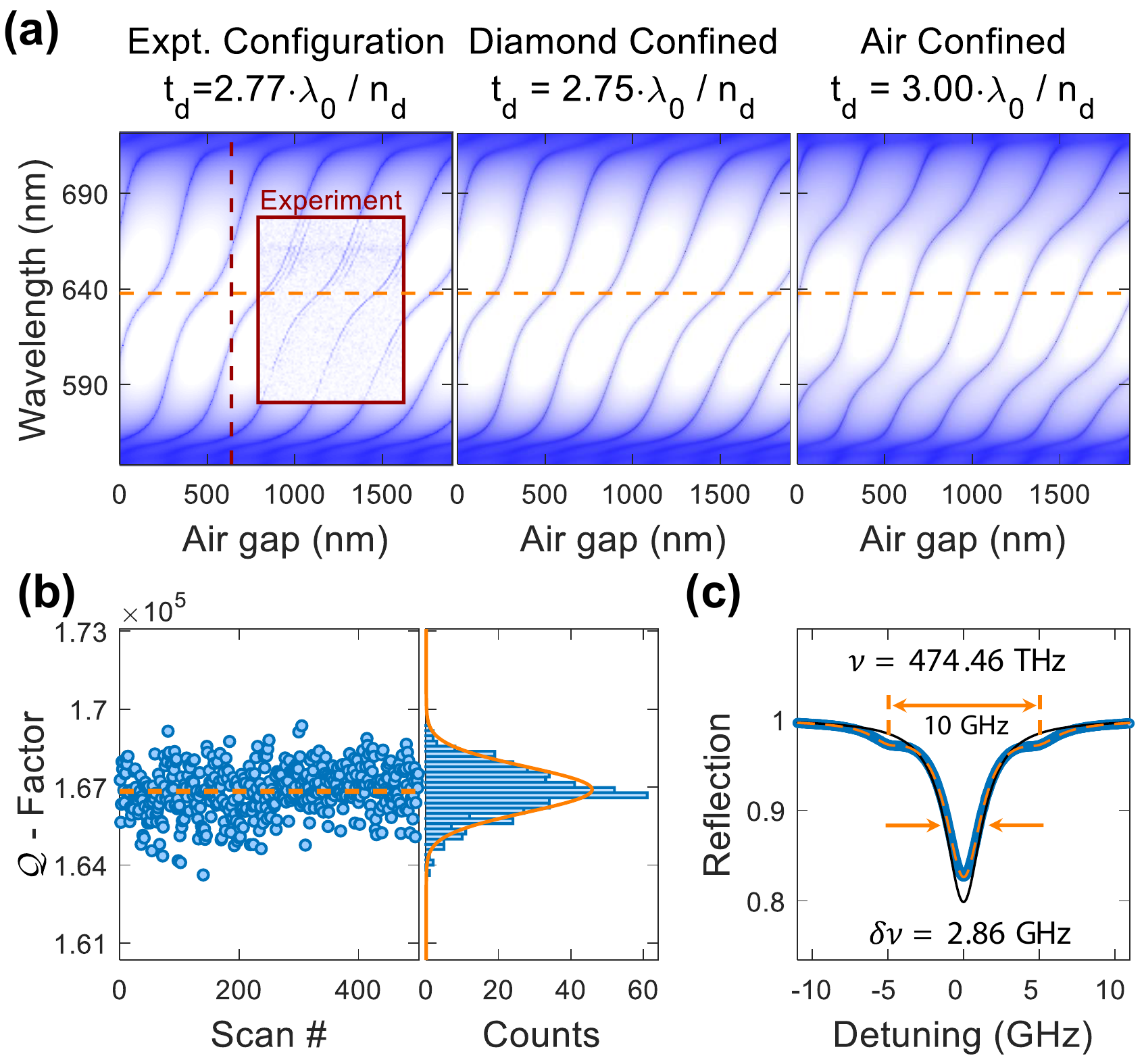}
\caption{\textbf{(a)} The inset in the leftmost panel shows PL as a function of cavity length under green excitation ($\lambda = 532\,\textrm{nm}$, $P = 30\,\textrm{mW}$). The non-linearity of the mode dispersion depends on the exact diamond-thickness. The experimental mode-structure (background) is well reproduced using a one-dimensional transfer-matrix simulation with $t_{\textrm{d}}=733\,\textrm{nm}$, corresponding to $t_{\textrm{d}}=2.77\frac{\lambda_0}{n_{\textrm{d}}}$ with $\lambda_0=637.7\,\textrm{nm}$. The vertical red dashed-line indicates the depth of the crater. The horizontal orange line indicates $\lambda_0=637.7\,\textrm{nm}$. The middle and rightmost panel show similar simulations for a diamond-confined  ($t_{\textrm{d}}=2.75\frac{\lambda_0}{n_{\textrm{d}}}=727\,\textrm{nm}$) and an air-confined ($t_{\textrm{d}}=3.00\frac{\lambda_0}{n_{\textrm{d}}}=794\,\textrm{nm}$) cavity, respectively. By comparison to the simulations, the geometry used in this experiment is clearly diamond-confined at the NV ZPL wavelength (orange dashed line, for details see main text). \textbf{(b)} Spread of 500 individual $\mathcal{Q}$-factor measurements on the diamond for mode $q_{\textrm{air}}=8$. The data follow a Gaussian distribution centred at $\mathcal{Q} = 166\,904$ with a standard deviation $\sigma = 874$. \textbf{(c)} Reflection of the cavity as a function of cavity detuning for $\lambda =631.9\,\textrm{nm}$. The blue data-points are the average of all the 500 scans displayed in  panel (b). The red line shows a triple Lorentzian-fit, where the side-peaks at $\nu_{\textrm{laser}}\pm 5\,\textrm{GHz}$ results from a frequency modulation which is employed as a frequency ruler. The black line is the reflected signal without any frequency modulation.}
\label{fig:ModeStruct}
\end{figure}

The mode structure exhibits two interesting features: a non-linear dispersion (an obvious feature in Fig.\,\ref{fig:ModeStruct}\,(a)) and the presence of higher-order transverse modes (weak feature in Fig.\,\ref{fig:ModeStruct}\,(a)). By analysing the spacing of the cavity modes (inset Fig.\,\ref{fig:ModeStruct}\,(a)) according to Eq.\,\ref{Effective_Cavity_Length}, we extract a radius-of-curvature $R_{\textrm{cav}}=21\,\upmu\textrm{m}$, in good agreement with the scanning confocal-microscope image shown in Fig.\,\ref{fig:intro}\,(a). We note that the detection optics were deliberately misaligned to facilitate detection of the higher-order modes (Fig.\,\ref{fig:ModeStruct}\,(a)).

The non-linear mode dispersion can be understood conceptually by considering a model consisting of two coupled cavities: one cavity-mode is confined to the diamond by the bottom DBR and the diamond-air interface; the other cavity mode is confined to the air-gap by the diamond-air interface and the top DBR. Across the diamond-air interface, these two cavity modes couple and hybridise, resulting in the avoided crossings depicted in the inset to Fig.\,\ref{fig:ModeStruct}\,(a)\,\cite{Janitz2015}. 

In this coupled diamond-air cavity model, the mode structure with changing air-gap $t_{\textrm{a}}$ and the position of the avoided crossings depend on the exact diamond-thickness $t_{\textrm{d}}$\,\cite{Janitz2015,Bogdanovic2017,Riedel2017,vanDam2018,Haussler2019,Riedel2020}. For a cavity of length $L=t_{\textrm{a}}+t_{\textrm{d}}$ (Fig.\,\ref{fig:Cav}), fundamental resonances occur provided $t_{\textrm{d}}n_{\textrm{d}}+t_{\textrm{a}}=j\frac{\lambda_0}{2}\, ,j \in \mathbb{N}$. Depending on $t_{\textrm{a,d}}$, two regimes emerge: the so-called ``diamond-confined" and ``air-confined" regimes\,\cite{Janitz2015}. For the diamond confined modes $t_{\textrm{d}}=(2i-1)\frac{\lambda_0}{4}\,, i \in \mathbb{N}$; for the air confined modes $t_{\textrm{d}}=i\frac{\lambda_0}{2}$\,\cite{RiedelPhD}. In the diamond confined geometry, a change in $t_{\textrm{a}}$ has a relatively small impact on the resonant wavelength, rendering the cavity robust against acoustic vibrations. A feature of the diamond-confined modes is that the vacuum electric-field amplitude is higher in the diamond than in the air-gap (Fig.\,\ref{fig:Cav}), leading to a relatively high coupling strength. However, an inevitable consequence of the diamond-confined modes is that the vacuum electric-field possesses an anti-node at the diamond-air interface\,\cite{Jensen2020}, thus exacerbating losses associated with scattering or absorption at the diamond-air interface\,\cite{Riedel2020}. Conversely,  in the air confined geometry, a small change in $t_{\textrm{a}}$ has a relatively large impact on the resonant wavelength, rendering the cavity sensitive to acoustic vibrations. A feature of the air-confined modes is that the vacuum electric-field is higher in the air-gap than in the diamond, reducing the coupling strength to an NV centre in the diamond\,\cite{RiedelPhD}. In this case, there is a node in the vacuum electric field at the diamond-air interface such that the design is relatively insensitive to losses at the diamond-air interface\,\cite{Riedel2020}.

Using a one-dimensional transfer-matrix simulation (Essential Macleod) we simulate the cavity mode-structure for different diamond thicknesses. We find an excellent agreement between the experiment (inset Fig.\,\ref{fig:ModeStruct}\,(a)) and the simulation (background Fig.\,\ref{fig:ModeStruct}\,(a)) for $t_{\textrm{d}}=733\,\textrm{nm}$. In this experiment, the width of the air-gap was made shorter and shorter until the two mirrors were in contact (at which point the resonant wavelengths of the cavity no longer depend on the applied piezo voltage). By considering the depth of the crater ($d\sim 0.64\,\upmu\textrm{m}$ from Fig.\,\ref{fig:intro}\,(a)), we extract a minimal mode number $q_{\textrm{air}}=3$ for the mode just out of contact. Here, $q_{\textrm{air}}$ is the mode index in air, starting at $q_{\textrm{air}}=1$ for the first resonance, corresponding to $t_{\textrm{a}}=129\,\textrm{nm}$ for $\lambda_0=637.7\,\textrm{nm}$. Both $q_{\textrm{air}}=1,2$ are inaccessible in this experiment on account of the depth of the top mirror-crater. The middle and rightmost panel in Fig.\,\ref{fig:ModeStruct} (a) show simulations for a diamond-confined ($t_{\textrm{d}}=2.75\frac{\lambda_{\textrm{0}}}{n_{\textrm{d}}}=727.4\,\textrm{nm}$) and for an air-confined ($t_{\textrm{d}}=3.00\frac{\lambda_{\textrm{0}}}{n_{\textrm{d}}}=793.5\,\textrm{nm}$) geometry, respectively. Here, $\lambda_{\textrm{0}}=637.7\,\textrm{nm}$ corresponds to the NV ZPL and $n_{\textrm{d}}$ is the refractive index of diamond. By comparing the experimental and simulated mode-structures it is clear that at the NV ZPL wavelength, the cavity operates in a diamond-confined regime.\\

The round-trip performance of the Fabry-Perot cavity is characterised by the finesse $\mathcal{F}$ defined as\,\cite{Hood2001,Hunger2010,Benedikter2015}
\begin{equation}
\mathcal{F}=\frac{2\pi}{\mathcal{L}_{\textrm{tot}}}\,, \label{Finesse}
\end{equation}
where $\mathcal{L}_{\textrm{tot}}=\mathcal{T}_{\textrm{t}}+\mathcal{T}_{\textrm{b}}+\mathcal{L}_{\textrm{cav}}$ is the fractional energy loss per round-trip. Here, $\mathcal{T}_{\textrm{t(b)}}$ is the transmission of the top (bottom) mirror and $\mathcal{L}_{\textrm{cav}}$ is the cavity round-trip-loss caused by additional loss mechanisms such as scattering or absorption. A reliable measurement of the finesse typically requires precise knowledge of the cavity linewidth over several free-spectral ranges (FSR). Such an experiment becomes impractical for high $\mathcal{F}$-values -- it requires a high dynamic-range. Conversely, a measurement of the $\mathcal{Q}$-factor, $\mathcal{Q}=\frac{\nu}{\delta\nu}$, requires knowledge only of the linewidth $\delta\nu$ for one cavity-mode only, a simpler experiment. For a cavity with perfect mirrors, the $\mathcal{Q}$-factor is linked to the finesse via
\begin{equation}
\mathcal{Q} = \frac{2L_{\textrm{cav}}}{\lambda}\mathcal{F}\,. \label{Q_and_Finesse}
\end{equation}

In the experiment, we tune the thickness of the air-gap $t_{\textrm{a}}$; $t_{\textrm{d}}$ remains constant. For fixed $\lambda$, provided the field penetrations into the mirrors remain constant, we write $L_{\textrm{cav}}=t_{\textrm{a}}+L_{\textrm{0}}$, where $t_{\textrm{a}}=q_{\textrm{air}}\frac{\lambda}{2}$. Here, $t_{\textrm{d}}$ and the field penetration into the mirrors are included in $L_{\textrm{0}}$. Thus, Eq.\,\ref{Q_and_Finesse} reduces to\,\cite{Kelkar2015}
\begin{equation}
\mathcal{Q}=q_{\textrm{air}}\mathcal{F}+\mathcal{Q}_{\textrm{0}}\, . \label{eq:Q_vs_q}
\end{equation}
 In other words, a measurement of the $\mathcal{Q}$-factor for subsequent modes ($q_{\textrm{air}}$ and $q_{\textrm{air}}+1$) determines the cavity finesse.

To determine the cavity linewidth $\delta\nu$, and thus the $\mathcal{Q}$-factor, we couple the output of a tunable diode-laser (Toptica DL Pro 635, $\lambda = 630\, ...\,640\,\textrm{nm}$ and  $\delta \nu \lesssim 500\,\textrm{kHz}$, $P=800\,\upmu\textrm{W}$) into the cavity. Keeping the excitation frequency $\nu_{\textrm{laser}}$ fixed, we tune the cavity length across the cavity resonance while monitoring the reflected signal using a photodiode and a fast oscilloscope (LeCroy WaveRunner 606Zi). To calibrate the piezo, and thus extract the cavity linewidth, we use an electro-optic modulator (EOM, Jenoptik PM635) to create laser side-bands at $\nu_{\textrm{laser}}\pm\,5\,\textrm{GHz}$\,\cite{Bogdanovic2017}. Here, we assume a linear behaviour of cavity length with piezo-voltage across the $10\,\textrm{GHz}$ bandwidth (corresponding to a change in air-gap, $\Delta t_{\textrm{a}}=0.056\,\textrm{nm}$). To extract reliably the cavity linewidth, the cavity is scanned across the resonance 500 times, each scan fitted independently with the sum of three Lorentzians. The $\mathcal{Q}$-factor is defined as the average value of all 500 scans. Fig.\,\ref{fig:ModeStruct}\,(b) shows the spread of the individually extracted $\mathcal{Q}$-factors for mode number $q_{\textrm{air}}=8$. Using a bin-size of $200$ for the values of $\mathcal{Q}$, the data follow a Gaussian centred around $\mathcal{Q} = 166\,900$ with standard deviation $\sigma = 870$. The blue line in Fig.\,\ref{fig:ModeStruct}\,(c) shows the average reflectivity data of all the 500 scans. Fitting a triple Lorentzian (orange line) yields an averaged cavity linewidth of $\delta\nu_{\textrm{avg}}=2.86\,\textrm{GHz}$, which gives $\mathcal{Q}_{\textrm{avg}}=165\,650$, in good agreement with the average of the individual scans. 

We present some details of the experiment. The linearly polarised red excitation-laser was passed through a $\lambda /2$-plate (B.\ Halle) before entering the cavity. A pellicle beam-splitter (Thorlabs BP145B1) was used to separate the reflected signal from the incident laser-beam. A linear polariser was used to isolate one of the two orthogonally-polarised cavity-modes in the reflected signal. (The mode-splitting arises either from a geometrical asymmetry of the curved mirror\,\cite{Foster2009,Uphoff2015} or from birefringence in the material comprising the cavity\,\cite{Hall2000,Brandstatter2013}). The cavity was scanned at a typical speed of $8.7\,\upmu\textrm{m/s}$ ($1.56\,\textrm{GHz/s}$). In the bare cavity, i.e.\ in a cavity without diamond membrane, for slow scanning speeds ($\lesssim 3\textrm{GHz/s}$) evidence of photothermal bistability\,\cite{Hunger2010,Brachmann2016} was observed. The origin of this effect is likely the weak absorption in the mirror coating on the order of $100\,\textrm{ppb}$\,\cite{An1997}. However, as these losses are negligible compared to the losses introduced by the diamond, the bistability was not investigated further. We note that photothermal bistability was not observed once the diamond membrane was included in the cavity.

\section{Results on $\mathcal{Q}$-factor}
\label{Sec:Results}
\begin{figure}[t]
\includegraphics[width=0.48\textwidth]{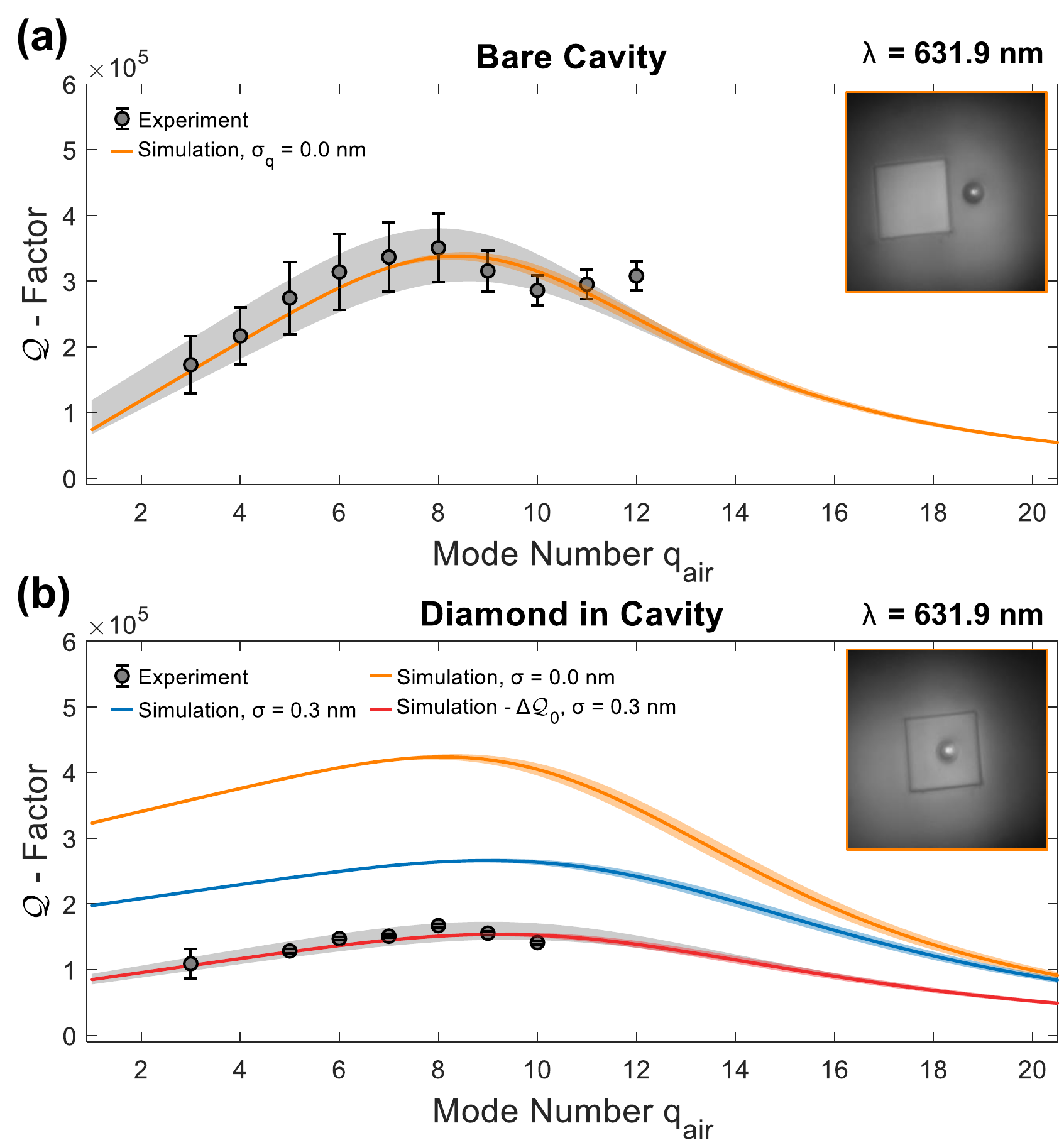}
\caption{\textbf{(a)} In black, the behaviour of the $\mathcal{Q}$-factor with increasing mode number $q_{\textrm{air}}$ for the bare cavity. The $\mathcal{Q}$-factor increases linearly for $q_{\textrm{air}}\leq 8$, after which clipping starts to occur. The orange line is the calculated $\mathcal{Q}$-factor using a 1D transfer-matrix model. \textbf{(b)} Introducing the diamond into the cavity reduces the $\mathcal{Q}$-factor (black data-points). Calculating the theoretical $\mathcal{Q}$-factor using a lossless model (orange) and scattering with surface roughness $\sigma_{\textrm{q}} = 0.3\,\textrm{nm}$ (blue) fail to reproduce the experimental values. The red line represents $\mathcal{Q}^{\textrm{sim}}-\Delta\mathcal{Q}_{\textrm{0}}$ with $\Delta\mathcal{Q}_{\textrm{0}}= 114\,000$, and describes the experiment well. For both panels, the black shaded regions account for the uncertainty in the fit parameters, while for the simulations, the shaded regions account for the uncertainty in the extracted tilt angle. For details see main text.}
\label{fig:Q_vs_q}
\end{figure}

In order to test our understanding of the mirrors themselves, we characterise initially the $\mathcal{Q}$-factor of the bare cavity, i.e.\ a cavity without a diamond membrane. Fig.\,\ref{fig:Q_vs_q}\,(a) shows the behaviour of the $\mathcal{Q}$-factor as a function of increasing mode number $q_{\textrm{air}}$ for fixed $\lambda = 631.9\,\textrm{nm}$. We observe a linear increase in $\mathcal{Q}$-factor for $q_{\textrm{air}}\leq 7$ as predicted by Eq.\,\ref{eq:Q_vs_q}. We attribute the drop in $\mathcal{Q}$-factor for $q_{\textrm{air}}>8$ to clipping losses at the top mirror\,\cite{Hunger2010}. Performing a linear fit for $q_{\textrm{air}}<8$ yields a bare-cavity finesse $\mathcal{F}_{\textrm{bare}}^{\textrm{exp}}=42\,500\pm4\,200$. The simulations predict $\mathcal{F}_{\textrm{bare}}^{\textrm{sim}}=44\,410$, in agreement with the experimental result to within the measurement uncertainty.

Next, we attempt to describe the dependence of the $\mathcal{Q}$-factor of the bare cavity on mode number $q_{\textrm{air}}$. Upon changing the cavity length $L$, the intensity beam  waist at the curved mirror $w_{\textrm{I}}$  evolves according to\,\cite{Nagourney}
\begin{equation}
w_{\textrm{I}}=\sqrt{\frac{\lambda R_{\textrm{cav}}}{\pi}}\left(\frac{R_{\textrm{cav}}}{L}-1\right)^{-\frac{1}{4}}\, .\label{eq:Beam_Waist}
\end{equation}
Clipping losses occur when this beam waist becomes larger than the spherical extent of the curved top-mirror\,\cite{Hunger2010,vanDam2018,Durak2014}. In principle, a small tilt-angle $\theta$ between the two mirrors will exacerbate clipping\,\cite{Flatten2016}. From a Gaussian optics approach\,\cite{SiegmanLasers}, we derive a model to estimate the clipping losses
\begin{equation}
\tilde{\mathcal{L}}_{\textrm{clip}}=e^{-\frac{D^2}{2w_{\textrm{I}}^2}}\left[1+\left(\frac{aD}{w_{\textrm{I}}^2}\right)^2\right]\,, \label{eq:ClippingLosses}
\end{equation}
where $a=R_{\textrm{cav}}\theta$ and $D$ is the diameter of the spherical extent of the mirror. In this model, the first term accounts for clipping\,\cite{Hunger2010,vanDam2018,SiegmanLasers} while the second term is a correction factor accounting for the tilt by angle $\theta$. In this model, the tilt results in a small lateral displacement of the cavity mode thereby increasing the clipping loss. Using the exact mirror-design obtained from Fig.\,\ref{fig:intro}\,(b), we simulate the behaviour of the cavity using a lossless 1D transfer-matrix simulation (Essential Macleod). The clipping losses are incorporated into the model according to $\mathcal{Q}_{\textrm{sim}}=\frac{4\pi L_{\textrm{cav}}}{\lambda}\left(\frac{1}{\mathcal{L}_{\textrm{sim}}+\mathcal{L}_{\textrm{clip}}}\right)$. To quantify the clipping losses, we perform a minimum mean-square error (MMSE) analysis, and find an excellent agreement using $D=5.9\,\upmu\textrm{m}$ and $\theta_{\textrm{bare}}=0\dots 0.27^{\circ}$. Including a 95\% confidence interval yields a maximum tilt-angle of $0.33^{\circ}$. The value of $D$ is in good accordance with the scanning confocal-image displayed in Fig.\ref{fig:intro}\,(a). The agreement between experiment and simulation indicates that intrinsic losses in the mirrors are negligibly small compared to losses introduced by the diamond, as discussed below.

Having established the intrinsic losses in the mirrors themselves, we introduce next the diamond membrane into the cavity by moving the bottom DBR in a lateral direction. Compared to the bare cavity, we observe a reduction in both $\mathcal{Q}$-factor and finesse (Fig.\,\ref{fig:Q_vs_q}\,(b), clear reduction in $\mathcal{Q}$-factor and $\mathcal{F}=\Delta\mathcal{Q}/\Delta q_{\textrm{air}}$ with respect to the simulation, respectively). Conceptually, the diamond effectively reduces the reflectivity of the bottom DBR, thus leading to a drop in the finesse (Eq.\,\ref{Finesse}). Performing a linear fit for $q_{\textrm{air}}<8$ yields $\mathcal{F}^{\textrm{exp}}_{\textrm{diamond}}=11\,500\pm 1\,100$. As before, we observe clipping for $q_{\textrm{air}}>8$. To quantify the clipping loss, we replace $L$ in Eq.\,\ref{eq:Beam_Waist} by $L_{\textrm{cav}}^{\textrm{d}}=t_{\textrm{a}}+\frac{t_{\textrm{d}}}{n_{\textrm{d}}}$\,\cite{vanDam2018} and apply Eq.\,\ref{eq:ClippingLosses} with $D=5.9\,\upmu\textrm{m}$. From a MMSE analysis, we calculate $\theta_{\textrm{d}} =\left( 0.37_{-0.26}^{+0.15}\right)^{\circ}$, where the high- and low limits are calculated from the 95\% confidence interval. The larger tilt angle might suggest a small thickness-gradient in the diamond membrane. 

Contrary to the bare-cavity case, a simulation using a lossless model (orange curve Fig.\,\ref{fig:Q_vs_q}\,(b)) fails to reproduce the experimental $\mathcal{Q}$-factors: the diamond membrane introduces additional loss mechanisms. Both the simulated $\mathcal{Q}$-factor and the finesse ($\mathcal{F}^{\textrm{sim}}_{\textrm{perfect}}=17\,450$) are larger than observed experimentally. We therefore need to introduce additional losses into our model. Working  in a diamond-confined regime, we expect these losses to occur at the diamond-air interface. 

We investigate the role of scattering at the diamond-air interface. To this end, we introduce a roughness of $\sigma_{\textrm{q}}=0.3\,\textrm{nm}$ at the diamond-air interface into the simulation\,\cite{Carniglia2002}. The choice of $\sigma_{\textrm{q}}$ is motivated by the AFM measurement displayed in Fig.\,\ref{fig:intro}\,(c) and from previously reported measurements\,\cite{Appel2016,Riedel2014,Riedel2017,Riedel2020}. The blue line in Fig.\,\ref{fig:Q_vs_q}\,(b) shows that scattering reduces both the $\mathcal{Q}$-factor and the finesse ($\mathcal{F}^{\textrm{sim}}_{\textrm{scat}}=10\,690$). Interestingly, we now observe that the simulated finesse, $\mathcal{F}^{\textrm{sim}}_{\textrm{scat}}$ is in good accordance with the experimentally determined finesse $\mathcal{F}^{\textrm{exp}}$ (equal $\Delta\mathcal{Q}/\Delta q_{\textrm{air}}$ in Fig.\,\ref{fig:Q_vs_q}\,(b)), while the simulated $\mathcal{Q}$-factor is larger than the experimentally determined value.
We rewrite Eq.\,\ref{eq:Q_vs_q} 
\begin{equation}
\mathcal{Q}^{\textrm{exp}}=\mathcal{Q}^{\textrm{sim}} - \Delta \mathcal{Q}_{\textrm{0}}\, .
\label{Q0}
\end{equation}
This pragmatic approach gives an accurate representation of the experiment (red line in Fig.\,\ref{fig:Q_vs_q}\,(b)) with $\Delta \mathcal{Q}_{\textrm{0}}=114\,000$.

We now aim to understand the origin of the losses introduced by the diamond, in particular the origin of the rigid reduction in $\mathcal{Q}$-factor described by the $\Delta\mathcal{Q}_{\textrm{0}}$-term. By measuring successive cavity modes for fixed $\lambda$, the beam waist at the bottom mirror evolves according to 
\begin{equation}
w_{\textrm{0,I}}=\sqrt{\frac{\lambda}{\pi}}\left(LR_{\textrm{cav}}-L^2\right)^{1/4}\,, \label{Eq:Beam_Waist_Mirror}
\end{equation}
 where $L=t_{\textrm{a}}+\frac{t_{\textrm{d}}}{n_{\textrm{d}}}$, thus probing a slightly larger surface area of the diamond\,\cite{Nagourney}. However, the standing-wave pattern at the diamond-air interface remains unaltered. Alternatively, changing the resonant $\lambda$ changes the standing wave inside the cavity. As scattering and absorption depend on the amplitude of the electric field, tuning the field maxima across the diamond-air interface may reveal the source of surface loss\,\cite{Najer2021}.

\begin{figure}[t]
\includegraphics[width=0.48\textwidth]{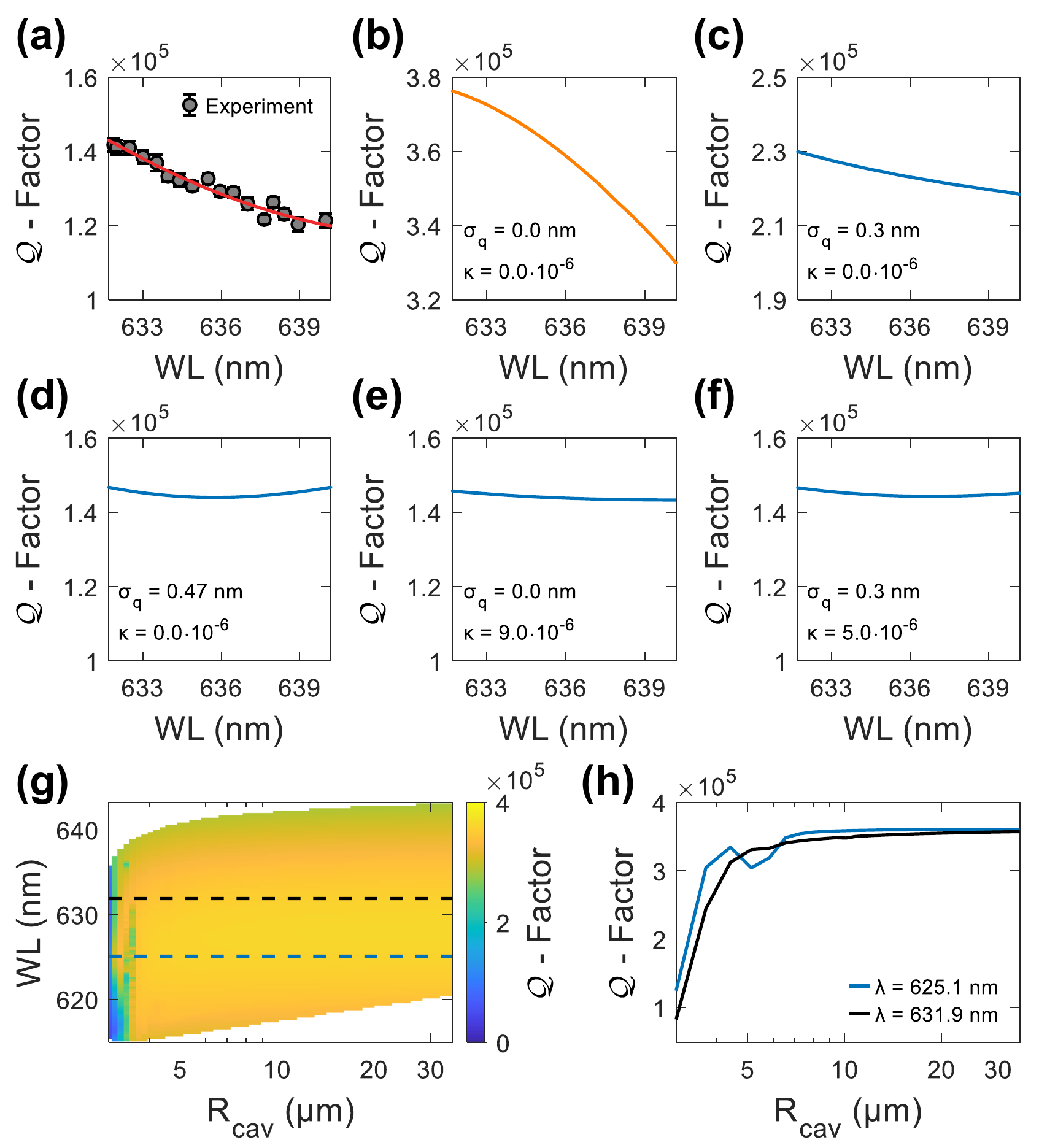}
\caption{\textbf{(a)} The measured $\mathcal{Q}$-factor as a function of wavelength for $q_{\textrm{air}}=4$. The $\mathcal{Q}$-factor drops for excitation wavelengths away from the stopband centre. The red line is a guide to the eye. \textbf{(b)} A calculation of the wavelength dependence of the $\mathcal{Q}$-factor for a lossless cavity. \textbf{(c)} Introducing scattering with surface roughness $\sigma_{\textrm{q}}=0.3\,\textrm{nm}$ reproduces the general behaviour of the experiment, but not the absolute numbers. \textbf{(d)}-\textbf{(f)} Calculations of the  $\mathcal{Q}$-factor with increased surface scattering ($\sigma_{\textrm{q}}$) and absorption ($\kappa$). \textbf{(g)} Calculated $\mathcal{Q}$-factor as function of wavelength and radius of curvature $R_{\textrm{cav}}$ for $q_{\textrm{air}}=4$. \textbf{(h)} The blue and black lines show the $\mathcal{Q}$-factor at the stopband centre ($\lambda_{\textrm{c}}=625\,\textrm{nm}$) and for $\lambda=631.9\,\textrm{nm}$, respectively. The significant drop in $\mathcal{Q}$-factor for $R_{\textrm{cav}}\lesssim 5-7\,\upmu\textrm{m}$ is attributed to clipping losses at the top mirror.
}
\label{fig:WL_dep}
\end{figure}

To this end, we measure the dependence of the $\mathcal{Q}$-factor on excitation wavelength $\lambda$ for mode $q_{\textrm{air}}=4$ (Fig.\,\ref{fig:WL_dep} (a)). We observe a drop in $\mathcal{Q}$-factor for wavelengths away from the stopband centre ($\lambda_{\textrm{c}}=625\,\textrm{nm}$, Fig.\,\ref{fig:intro}\,(b)). As before, a lossless model (Fig.\,\ref{fig:WL_dep}\,(b)) fails to reproduce the absolute value of the $\mathcal{Q}$-factor as well as the dependence on $\lambda$.

We consider enhanced diamond-related losses, surface scattering and absorption in the diamond itself, as the origin of $\Delta\mathcal{Q}_{\textrm{0}}$. In Fig.\,\ref{fig:WL_dep}\,(d) we increase the surface roughness to $\sigma_{\textrm{q}}=0.47\,\textrm{nm}$. Next, in Fig.\,\ref{fig:WL_dep}\,(e)  we include absorption in the diamond by varying the value of the extinction coefficient $\kappa$\,\cite{Carniglia2002}. Finally, in Fig.\,\ref{fig:WL_dep}\,(f) we combine surface scattering ($\sigma_{\textrm{q}}=0.3\,\textrm{nm}$) with absorption ($\kappa=5.6\cdot 10^{-6}$). All three simulations accurately account for the $\mathcal{Q}$-factor at short $\lambda$. However, the simulations fail to reproduce the behaviour with increasing $\lambda$. The simulations predict a minimum $\mathcal{Q}$-factor for $\lambda\sim636\,\textrm{nm}$ beyond which an increase in $\mathcal{Q}$-factor is predicted, a feature not observed experimentally. It would appear therefore that a combination of surface roughness and absorption cannot be responsible for $\Delta\mathcal{Q}_{\textrm{0}}$. Furthermore, significant absorption in the diamond is unlikely -- it results in a weak dependence of the $\mathcal{Q}$-factor on wavelength, yet in the experiments there is a strong wavelength dependence.

Another factor to consider are diffraction losses. Up until this point, only one-dimensional transfer-matrix simulations were performed; these simulations do not consider any diffraction loss at the top DBR. In addition, for tightly confined modes, the angular spread in $k$-space increases, leading to an increased loss in the DBR mirror and thus a reduction in $\mathcal{Q}$-factor\,\cite{Tomm2021}. To investigate this, we perform numerical simulations (COMSOL Multiphysics) of the $\mathcal{Q}$-factor as a function of $R_{\textrm{cav}}$ and $\lambda$ (Fig.\,\ref{fig:WL_dep}\,(g)). Looking at a linecut for fixed $\lambda$ (Fig.\,\ref{fig:WL_dep}\,(h)), we observe a strong dependence of $\mathcal{Q}$-factor with radius for $R_{\textrm{cav}}\lesssim5-7\,\upmu\textrm{m}$. For larger radii, this dependence is weak, and the $\mathcal{Q}$-factor saturates at $\mathcal{Q}=360\,000$ in good agreement with the one-dimensional transfer-matrix simulations. We therefore conclude that diffraction losses at the top mirror are negligible, that the one-dimensional simulations provide reliable predictions even of the behaviour of the three-dimensional cavity, and that diffraction is not responsible for $\Delta\mathcal{Q}_{\textrm{0}}$.

Based on this understanding, we simulate the cavity $\mathcal{Q}$-factor by including a scattering layer at the diamond-air interface with $\sigma_{\textrm{q}}=0.3\,\textrm{nm}$ (Fig.\,\ref{fig:WL_dep}\,(c)), taking the absorption in the diamond and likewise any diffraction losses to be negligibly small. This approach reproduces the experimentally observed decrease of the $\mathcal{Q}$-factor with $\lambda$.

This analysis suggests that close to the stopband centre, scattering at the diamond-air interface reduces the $\mathcal{Q}$-factor from an ideal value of $375\,540$ to $229\,330$. An additional loss mechanism, which results in the $\Delta\mathcal{Q}_{\textrm{0}}$-term, reduces the $\mathcal{Q}$-factor further to a value of $141\,100$.  We note that if we assume that the experimental finesse matches the simulated finesse at all wavelengths then $\Delta\mathcal{Q}_0$ has a small wavelength dependence, increasing monotonically by about 15\% from $\lambda=630\,\textrm{nm}$ to $\lambda=640\,\textrm{nm}$. The microscopic origin of the $\Delta\mathcal{Q}_{\textrm{0}}$-term is not known precisely. We speculate that it arises as a consequence of the waviness in the profile of the diamond surface (Fig.\,\ref{fig:intro}\,(c)). The spatial frequency of the waviness is comparable to that of the cavity mode -- the waviness does not scatter in the same way as the surface roughness. Compatible with this hypothesis is the observation that the $\mathcal{Q}$-factor is position dependent: the measured $\mathcal{Q}$-factor was rather low at certain locations of the diamond membrane. It is an open question how the waviness might result in a rigid reduction of the $\mathcal{Q}$-factor according to Eq.\,\ref{Q0}.
\section{Prediction on the Purcell factor}
Improvements in the optical properties of an NV centre in a resonant microcavity depend on the Purcell factor\,\cite{Purcell1946}. Based on the experimental results, we investigate the potential Purcell factors in a cavity of this type. To do this, we make the assumptions that better fabrication can eliminate the losses implied in the $\Delta\mathcal{Q}_{\textrm{0}}$-term; that the surface roughness of $\sigma_{\textrm{q}}=0.3\,\textrm{nm}$ is already excellent -- some surface scattering is therefore inevitable; and that the absorption losses in the diamond are negligible; and that we work with the mirrors from the experiment with their slight imperfections.

\begin{figure}
\includegraphics[width=0.48\textwidth]{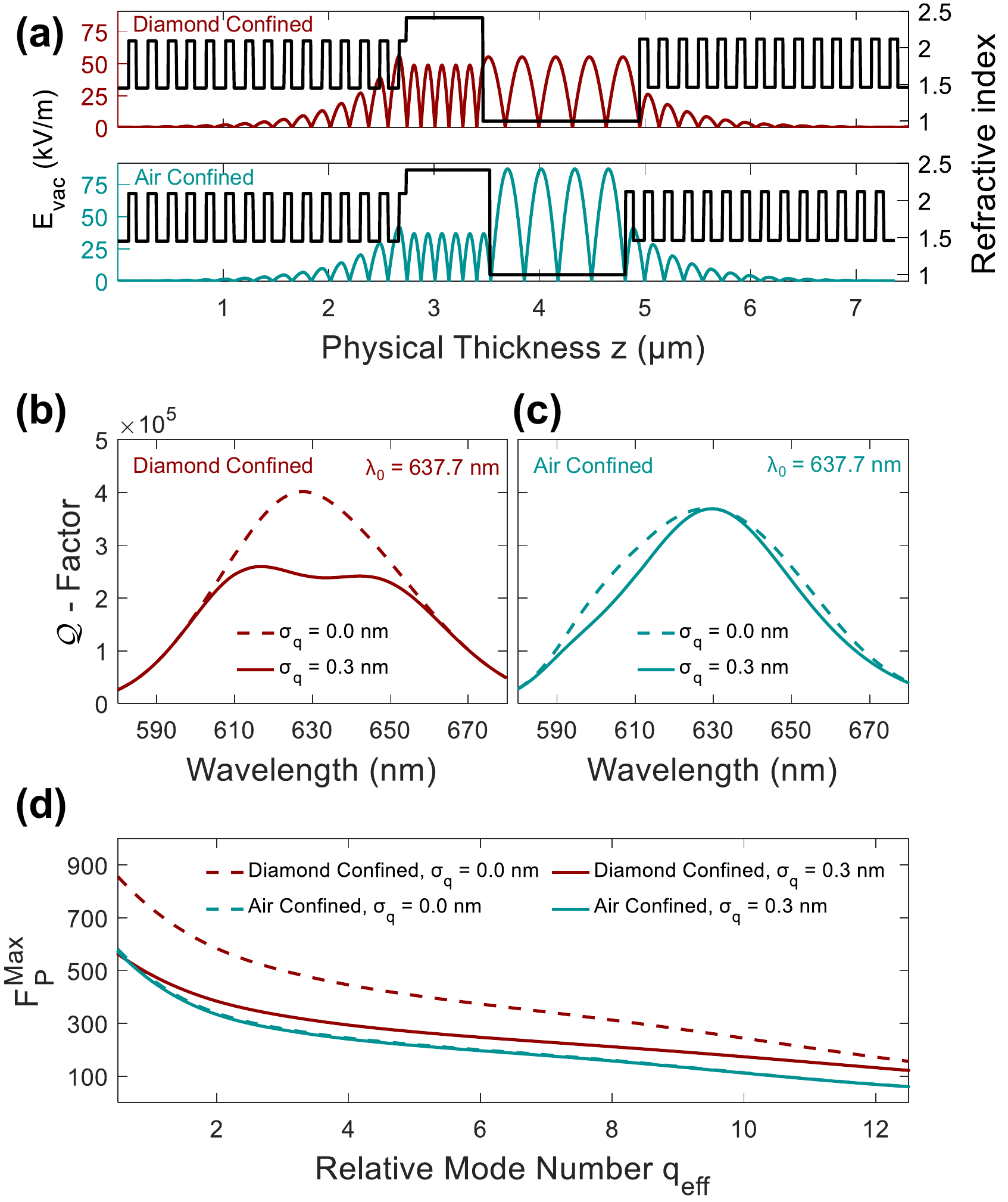}
\caption{\textbf{(a)} The vacuum electric-field distribution for a diamond-confined (top, $t_{\textrm{d}}=2.75\frac{\lambda_0}{n_{\textrm{d}}}=727\,\textrm{nm}$) and air-confined (bottom, $t_{\textrm{d}}=3.00\frac{\lambda_0}{n_{\textrm{d}}}=794\,\textrm{nm}$) geometry obtained from a one-dimensional transfer-matrix simulation using the mirror design extracted from Fig.\,\ref{fig:intro}\,(b). The diamond-confined case exhibits a field anti-node at the diamond-air interface, while the air-confined geometry exhibits a field node at the diamond-air interface. \textbf{(b)}-\textbf{(c)} Simulation of the $\mathcal{Q}$-factor as a function of wavelength for diamond-confined (b) and air-confined (c) geometries. Introducing surface scattering with $\sigma_{\textrm{q}}=0.3\,\textrm{nm}$ reduces the $\mathcal{Q}$-factor in the diamond-confined case, while for the air-confined geometry, the $\mathcal{Q}$-factor remains relatively unaltered. \textbf{(d)} Expected Purcell factor as a function of mode number $q_{\textrm{air}}$.
}
\label{fig:Theory}
\end{figure}

To understand fully the behaviour of $\mathcal{Q}$ and with $\lambda$, we need to consider the standing wave inside the cavity. Fig.\,\ref{fig:Theory}\,(a) shows the profile of the vacuum electric-field for a diamond-confined ($t_{\textrm{d}}=2.75\frac{\lambda_{\textrm{0}}}{n_{\textrm{d}}}$) and air-confined ($t_{\textrm{d}}=3.00\frac{\lambda_{\textrm{0}}}{n_{\textrm{d}}}$) cavity, respectively. Here, $\lambda_{\textrm{0}}=637.7\,\textrm{nm}$. For the diamond-confined geometry, there is a field maximum at the diamond-air interface. Surface scattering scales with the amplitude of the electric field, thus, for $\lambda=637.7\, \textrm{nm}$ scattering is maximised resulting in a minimum $\mathcal{Q}$-factor. For $\lambda$ away from 637\,nm, the field amplitude goes down, thus the losses are reduced and the $\mathcal{Q}$-factor goes up. Fig.\,\ref{fig:Theory}\,(b) and (c) show the calculated behaviour of the $\mathcal{Q}$-factor over a large range of $\lambda$ for a diamond- and air-confined geometry, respectively. Introducing scattering reduces the $\mathcal{Q}$-factor significantly for the diamond-confined geometry, while for the air-confined geometry, the $\mathcal{Q}$-factor remains relatively unaltered.

We now calculate the expected Purcell factor\,\cite{Purcell1946} for our device. To start, we simulate the vacuum electric-field distribution for a one-dimensional cavity using the same transfer-matrix algorithm used to simulate the $\mathcal{Q}$-factor (Essential Macleod). For a Gaussian cavity-mode, the vacuum electric-field is quantised according to\,\cite{Riedel2020}
\begin{equation}
\begin{split}
\int_{\rm cav}\epsilon_0 \epsilon_\text{R}(z) |\vec{E}_\text{vac}(z)|^2 {\rm d}z\int_0^{2\pi}{\rm d}\phi\int_0^{\infty} r e^{-r^2/2 w_\text{I}^2}{\rm d}r\\
=2 \pi \frac{1}{4}w_{\textrm{I}}^2 \int_{\rm cav}\epsilon_0 n^2(z) |\vec{E}_\text{vac}(z)|^2 {\rm d}z=\frac{\hbar\omega}{2}\,. \label{Eq:Evac}
\end{split}
\end{equation}
Here, we take $\epsilon_{\textrm{R}}=n_{\textrm{d}}^2$ and assume a constant beam waist $w_{\text{0,d}}\simeq 1.0\,\upmu\textrm{m}$ ($q_{\textrm{air}}=4$) along the length of the cavity, calculated from Eq.\,\ref{Eq:Beam_Waist_Mirror}. Inside the diamond, we obtain a maximum $\lvert \vec{E}_{\textrm{vac}}\rvert = 54.73\,\textrm{kVm}^{-1}$. For an emitter located at $\vec{r}=\vec{r}_{\textrm{0}}$, the effective mode-volume is calculated according to\,\cite{Santori2010,Kristensen2012}
\begin{equation}
\begin{split}
V_{\textrm{eff}}=\frac{\int_{\textrm{cav}}\epsilon_{\textrm{0}}\epsilon_{\textrm{R}}(\vec{r})\lvert\vec{E}_{\textrm{vac}}(\vec{r})\rvert^2\textrm{d}^3 r}{\epsilon_{\textrm{0}}\epsilon_{\textrm{R}}(\vec{r}_{\textrm{0}})\lvert\vec{E}_{\textrm{vac}}(\vec{r}_{\textrm{0}})\rvert^2} \\
=\frac{\hbar\omega/2}{\epsilon_{\textrm{0}}\epsilon_{\textrm{R}}(\vec{r}_{\textrm{0}})\lvert\vec{E}_{\textrm{vac}}(\vec{r}_{\textrm{0}})\rvert^2}\, . \label{eq:modeVolume}
\end{split}
\end{equation}
Numerically, we obtain $V_{\textrm{eff}} = 54.17\left(\frac{\lambda}{n}\right)^3$. For the experimental geometry, $\mathcal{Q}^{\textrm{sim}}_{\sigma_{\textrm{q}}=0.3\,\textrm{nm}}=221\,000$ for $\lambda =637.7\,\textrm{nm}$, from which we deduce
\begin{equation}
F_{\textrm{P}}=1+\frac{3}{4\pi^2}\frac{\mathcal{Q}^{\textrm{sim}}_{\sigma_{\textrm{q}}=0.3\,\textrm{nm}}}{V_{\textrm{eff}}}\left(\frac{\lambda}{n}\right)^3=309\,. \label{Eq:Purcell}
\end{equation}
The probability of emission into the cavity mode for an emitter with 100\,\% quantum efficiency is given by the $\beta$-factor: $\beta=\frac{F_{\textrm{P}}-1}{F_{\textrm{P}}}=0.9968$. We note that the Purcell factor is independent of any emitter properties: the calculation is based solely on the experimental cavity parameters\,\cite{Janitz2020}.

Next, we apply the calculated Purcell-factor to an NV centre: we are interested in calculating the emission rate into the ZPL. We assume that the NV centre optical dipole is aligned along the polarisation-axis of the cavity mode. In the absence of the cavity, the excited-state decay-rate $\gamma_{\textrm{free}}=\xi_{\textrm{0}}\gamma_{\textrm{0}}+(1-\xi_{\textrm{0}})\gamma_{\textrm{0}}$, where $\xi_{\textrm{0}}$ is the Debye-Waller factor describing the branching ratio into the ZPL and $\gamma_{\textrm{0}}=\frac{1}{\tau_{0}}$ with $\tau_{\textrm{0}}$ the radiative lifetime. Here, we have assumed unity internal quantum efficiency ignoring any non-radiative decay channels\,\cite{Riedel2017,Ruf2021}. Tuning the cavity on resonance with the ZPL enhances the ZPL emission by $F_{\textrm{P}}$ while the emission into the phonon-sideband remains unaltered. Therefore, in the presence of the cavity, the decay rate becomes $\gamma_{\textrm{cav}}=F_{\textrm{P}}\xi_{\textrm{0}}\gamma_{\textrm{0}}+(1-\xi_{\textrm{0}})\gamma_{\textrm{0}}$, where $F_{\textrm{P}}$ is defined according to Eq.\,\ref{Eq:Purcell}\,\cite{Ruf2021}. Taking the ratio of the decay rate in the cavity to that of free space yields
\begin{equation}
\frac{\gamma_{\textrm{cav}}}{\gamma_{\textrm{free}}}=\frac{\tau_{\textrm{0}}}{\tau_{\textrm{cav}}}=1+\xi_{\textrm{0}}\left(F_{\textrm{P}}-1\right)\, , \label{eq:Ratio_Decay_Rate}
\end{equation}
where $\tau_{\textrm{cav}}$ is the radiative lifetime in the cavity. Taking the unperturbed lifetime $\tau_{\textrm{0}}=12.6\,\textrm{ns}$ and ${\xi_{\textrm{0}}=2.55\%}
$\,\cite{Riedel2017} along with $F_{\textrm{P}}=309$, Eq.\,\ref{eq:Ratio_Decay_Rate} predicts an reduction in lifetime to $\tau_{\textrm{cav}}=1.42\,\textrm{ns}$. The reduction in lifetime results in a broadening of the homogeneous linewidth from $\Delta\nu_{\textrm{free}}=\frac{1}{2\pi}\gamma_{\textrm{0}}=12.6\,\textrm{MHz}$ to ${\Delta\nu_{\textrm{cav}}^{\textrm{ZPL}}=\frac{1}{2\pi}\left[1+\xi_{\textrm{0}}\left(F_{\textrm{P}}-1\right)\right]\gamma_{\textrm{0}}=112\,\textrm{MHz}}$, rendering the NV less sensitive to spectral wandering. Finally, we calculate the efficiency, $\eta_{\textrm{ZPL}}$, of emitting a photon into the ZPL\,\cite{Ruf2021}; ${\eta_{\textrm{ZPL}}=F_{\textrm{P}}\frac{\xi_{\textrm{{0}}}\gamma_{\textrm{0}}}{\gamma_{\textrm{cav}}}=\frac{\xi_{\textrm{0}}F_{\textrm{P}}}{\xi_{\textrm{0}}\left(F_{\textrm{P}}-1\right)+1}=89.0\,\%}$ 

Alternatively, the NV-cavity coupling can be described with the Jaynes-Cummings Hamiltonian in terms of $\{g_{\textrm{ZPL}},\kappa,\gamma_{0}\}$: where $g_{\textrm{ZPL}}=d_{\textrm{NV}}E_{\textrm{vac}}$ is the NV-cavity coupling rate, $\kappa$ is the cavity decay rate and $\gamma_0$ is, as before, the spontaneous emission rate\,\cite{Reiserer2015,Barrett2020}. Using $d_{\textrm{NV}}/e=\sqrt{\xi_0}0.108\,\textrm{nm}$\,\cite{Riedel2017}, we deduce $\{g_{\textrm{ZPL}},\kappa,\gamma_0\}=2\pi\times\{228\,\textrm{MHz},\,2.13\,\textrm{GHz},\,12.63\,\textrm{MHz}\}$, firmly placing the system in the weak-coupling regime of cavity QED. The condition $(\kappa>g>\gamma)$  is favourable for photon collection\,\cite{Ruf2021}. This approach results in ${\eta_{\textrm{ZPL}}=\frac{4g_{\textrm{ZPL}}^2/(\kappa\gamma_0)}{4g_{\textrm{ZPL}}^2/(\kappa\gamma_0)+1}=88.6\,\%}$, and gives the same numerical value as above.

We now compare the potential Purcell factors for diamond-confined and air-confined cavities. There is a trade-off: the diamond-confined cavity has a larger $E_{\textrm{vac}}$ at the location of an optimally-positioned NV centre but is more sensitive to scattering at the diamond-air surface with respect to the air-confined cavity. Fig.\,\ref{fig:Theory}\,(d) shows a comparison between the Purcell factor for a diamond-confined and air-confined cavity ($t_{\textrm{d}}=2.75\lambda_{\textrm{0}}$ and $t_{\textrm{d}}=3.00\lambda_{\textrm{0}}$, respectively). In the absence of any surface losses, the Purcell factor is significantly larger for the diamond-confined geometry compared to an air-confined geometry owing to two factors: the larger effective-length yields a higher $\mathcal{Q}$-factor, and the stronger confinement of the vacuum field to the diamond yields a lower effective mode volume. However, introducing surface scattering ($\sigma_{\textrm{q}}=0.3\,\textrm{nm}$ as before) reduces the Purcell factor for the diamond-confined geometry, while for the air-confined geometry the Purcell factor remains roughly the same. Despite the higher losses associated with a surface roughness of $\sigma_{\textrm{q}}=0.3\,\textrm{nm}$, the calculations suggest that it is beneficial to work in a diamond-confined geometry on account of the higher Purcell factor (at e.g.\ $q_{\textrm{air}}=4$, Fig.\, \ref{fig:Theory}\,(d)) -- this will result in a higher flux of coherent photons. An additional benefit of practical importance is that for the diamond-confined geometry $\textrm{d}\lambda\,/\textrm{d}t_{\textrm{a}}=0.11$ compared to $\textrm{d}\lambda\,/\textrm{d}t_{\textrm{a}}=0.27$ for the air-confined geometry with same mode-number $q_{\textrm{air}}$ (Fig.\,\ref{fig:ModeStruct}\,(a)), thus rendering the cavity less susceptible to acoustic vibration\,\cite{Riedel2020}.

\section{Conclusions and outlook}
In this work, we have demonstrated the possibility of achieving high $\mathcal{Q}$-factors in a Fabry-Perot resonator in which the electromagnetic field is strongly confined to a diamond membrane. A $\mathcal{Q}$-factor of $121\,700$ was achieved for $\lambda\simeq637\,\textrm{nm}$ for the minimum mode number, $q_{\textrm{air}}=4$. The $\mathcal{Q}$-factor is lower than the $\mathcal{Q}$-factor expected from the geometry alone. The main source of loss in this experiment is attributed to roughness and waviness at the diamond surface. The waviness, attributed as polishing marks, can potentially be mitigated by optmised plasma etching\,\cite{Appel2016} and/or by atomic-layer deposition of a material with refractive index less than diamond\,\cite{vanDam2018}. Deposition of SiO$_2$ ($n=1.47$) or Al$_2$O$_3$ ($n=1.77$) will also reduce the losses due to scattering. We note that surface passivation has previously been demonstrated to increase the $\mathcal{Q}$-factor for GaAs resonators\,\cite{Guha2017,Najer2019,Najer2021} albeit via a different mechanism.

Despite the presence of surface-related losses, the current design is capable of reaching a theoretical Purcell factor $F_{\textrm{P}}=170$. If the waviness can be eliminated leaving the surface roughness the same, the current design is capable of reaching $F_{\textrm{P}}=309$. Without the surface waviness but with the existing surface roughness, the Purcell factor is predicted to be higher for a diamond-confined cavity with respect to an air-confined cavity. 

The motivation behind this work is to enhance the flux of coherent photons from single NV centres in diamond\,\cite{Riedel2017}, a step towards the realisation of an efficient spin-photon interface\,\cite{Awschalom2018}. We note that the Purcell factor presented here is universal: $F_{\textrm{P}}$ depends solely on the cavity parameters, not on the properties of the emitter. The versatile design of the cavity allows a wide-range of solid-state single-photon emitters to be embedded\,\cite{Aharonovich2016}, for instance other colour centres in diamond\,\cite{Aharonovich2011,Iwasaki2017,Rose2018,Bradac2019,Trusheim2019,Baier2020,Chen2020,Rugar2021}, defects in SiC\,\cite{Riedel2012,Economou2016,Christle2017,Anderson2019,Lukin2020}, rare-earth ions in a crystalline host\,\cite{Zhong2017,Casabone2018,Zhong2019,Kindem2020,Chen2020Science} or emitters in 2D materials\,\cite{Caldwell2019,Dietrich2020}.

\section*{Acknowledgements}
We thank Viktoria Yurgens for fruitful discussions and Lukas Sponfeldner for help with the atomic-force microscopy measurements. We acknowledge financial support from National Centre of Competence in Research (NCCR) Quantum Science and Technology (QSIT), a competence center funded by Swiss National Science Foundation (SNF), the Swiss Nanoscience Institute (SNI), Innovative Training Network (ITN) network SpinNANO, the European Union (EU) Quantum Flagship project ASTERIQS grant agreement No.\ 820394. D.R. acknowledges support from the SNF (Project P400P2\_194424). T.J. acknowledges support from the European Unions Horizon 2020 Research and Innovation Programme under the Marie Sk\l{}odowska-Curie grant agreement No.\ 792853 (Hi-FrED) and support from the Polish National Agency for Academic Exchange under Polish Returns 2019 programme (agreement PPN/PPO/2019/1/00045/U/0001). A.J. acknowledges support from the European Unions Horizon 2020 Research and Innovation Programme under Marie Sk\l{}odowska-Curie grant agreement no.\ 840453 (HiFig).

%

\end{document}